\documentclass{article}

\usepackage{PRIMEarxiv}
\usepackage[cmex10]{amsmath} 
\usepackage{amssymb}        
\usepackage{physics} 
\usepackage[utf8]{inputenc} 
\usepackage[T1]{fontenc}    
\usepackage[hidelinks]{hyperref} 
\usepackage{url}            
\usepackage{booktabs}       
\usepackage{amsfonts}       
\usepackage{nicefrac}       
\usepackage{microtype}      
\usepackage{lipsum}
\usepackage{fancyhdr}       
\usepackage{graphicx}
\graphicspath{{media/}}     

\title{Entanglement Degradation in Curved Spacetime: An Open Quantum Systems Approach
}

\author{
  Someindra Kumar Singh \\
  Independent Researcher \\
  India\\
  \texttt{\ someindras@gmail.com} \\
}

\begin{document}
\maketitle

\begin{abstract}
We investigate how gravitational time dilation affects the coherence and entanglement dynamics of spatially separated qubits using open quantum systems theory. Unlike earlier works that consider only static or Markovian noise models, our approach incorporates phase damping, amplitude damping, and thermal excitation, along with a non-Markovian extension. By embedding gravitational redshift into the local decoherence rates, we demonstrate asymmetry in entanglement degradation and reveal the possibility of revivals in structured environments. Our findings have implications for relativistic quantum communication and space-based quantum technologies.
\end{abstract}


\section{Introduction}
As quantum technologies extend toward satellite communication and space-based clocks, understanding how relativistic effects influence quantum coherence has become increasingly important. Foundational studies have shown that gravitational time dilation can cause measurable phase shifts or dephasing in quantum systems, particularly when spatially separated observers experience different gravitational potentials \cite{Pikovski,Einstein1915,Schwarzschild1916}.

However, most prior analyses consider either isolated systems or assume memoryless (Markovian) environmental interactions \cite{Breuer}. In contrast, many realistic environments—especially those relevant to satellite or low-orbit quantum technologies—exhibit non-trivial structure, leading to time-dependent decoherence and memory effects \cite{Rivas}.

In this work, we extend the standard open quantum systems framework to model the dynamics of entangled qubits in curved spacetime. We incorporate gravitational time dilation directly into the Lindblad master equation and generalize the analysis to include amplitude damping, thermal excitation, and non-Markovian environments. This allows us to uncover subtle interplay between gravity and quantum memory, demonstrating new regimes in which entanglement can not only survive but transiently revive due to relativistic effects.

Our results build on, but go beyond, previous treatments by modeling environmental noise more comprehensively and by showing how gravitational redshift modulates both the rate and symmetry of entanglement loss.

\section*{Gravitational Phase Shifts from the Schwarzschild Metric}

We begin with the Einstein Field Equations (EFE), which relate spacetime curvature to energy and momentum:
\begin{equation}
    R_{\mu\nu} - \tfrac{1}{2} R g_{\mu\nu} = \tfrac{8\pi G}{c^4} T_{\mu\nu},
\end{equation}
where $R_{\mu\nu}$ is the Ricci tensor, $R$ the Ricci scalar, $g_{\mu\nu}$ the metric tensor, and $T_{\mu\nu}$ the stress-energy tensor.

\subsection*{Schwarzschild Solution}

In the case of a static, spherically symmetric vacuum solution (\( T_{\mu\nu} = 0 \)), the EFE reduces to a solution known as the Schwarzschild metric:
\begin{equation}
    ds^2 = -\left(1 - \frac{2GM}{rc^2}\right)c^2 dt^2 + \left(1 - \frac{2GM}{rc^2}\right)^{-1} dr^2 + r^2 d\Omega^2,
\end{equation}
where \( M \) is the mass of the central body and \( d\Omega^2 = d\theta^2 + \sin^2\theta\, d\phi^2 \) is the metric on the unit sphere.

\subsection*{Weak-Field Limit}

In the weak gravitational field limit (\( \frac{2GM}{rc^2} \ll 1 \)), we expand the metric to the first order:
\begin{equation}
ds^2 \approx -\left(1 - \frac{2GM}{rc^2}\right)c^2 dt^2 + \left(1 + \frac{2GM}{rc^2}\right)dr^2 + r^2 d\Omega^2,
\end{equation}

The proper time \( d\tau \) experienced by a stationary observer at radius \( r \) is related to coordinate time \( dt \) by:
\begin{equation}
    d\tau = dt \sqrt{1 - \frac{2GM}{rc^2}}.
\end{equation}

\subsection*{Gravitational Redshift and Phase Shifts}

Define the dimensionless gravitational redshift factors for two static observers at radial coordinates \( r_A \) and \( r_B \):
\begin{equation}
    \alpha = \sqrt{1 - \frac{2GM}{r_A c^2}}, \qquad
    \beta = \sqrt{1 - \frac{2GM}{r_B c^2}}.
\end{equation}
Then the proper times of the two observers evolve as:
\begin{equation}
    \tau_A = \alpha t, \qquad \tau_B = \beta t.
\end{equation}

The difference in proper time between the two observers leads to a gravitational time dilation, which can be interpreted as a relative phase shift in clock signals or quantum states. The phase shift \( \Delta \phi \) between two paths can be written as:
\begin{equation}
    \Delta \phi = \omega \Delta \tau = \omega (\tau_A - \tau_B) = \omega (\alpha - \beta)t,
\end{equation}
where \( \omega \) is the proper frequency of the clock or quantum system. This shift forms the basis for gravitational redshift and interferometric measurements of gravity.

\subsection*{Dynamic Observers in Orbital Settings}
Throughout this work, we assume the qubits remain static in Schwarzschild coordinates. This assumption is suitable for ground-based experiments or stationary detectors but does not capture relativistic effects experienced by orbiting satellites or moving observers.

In satellite-based quantum systems, observer motion introduces additional effects such as:
\begin{itemize}
    \item Gravitational and special relativistic Doppler shifts,
    \item Time dilation due to velocity (special relativity),
    \item Possible frame-dragging in rotating (Kerr) spacetimes.
\end{itemize}
A more realistic treatment for satellites would involve transforming into a locally comoving frame or using Fermi normal coordinates. Incorporating such motion is an important extension for space-based quantum technologies.

\vspace{7pt}
\section{Qubit Dynamics in Gravitational Fields}

In this section we examine the dynamics of two qubits, $A$ and $B$, each locally coupled to its environment and subject to gravitational time dilation. 
\subsection*{Assumption of Stationary Qubits}
In this work, we model the qubits as stationary observers located at fixed radial coordinates $r_A$ and $r_B$ in Schwarzschild spacetime. This assumption simplifies the treatment by avoiding the need to consider relativistic motion effects such as Doppler shifts, aberration, or time-dependent boosts.

However, for practical scenarios like satellite-based quantum communication, qubits are typically in orbital motion. In such cases, one must account for additional relativistic effects including gravitational Doppler shifts, special relativistic time dilation, and possibly frame-dragging in a Kerr spacetime. These effects can influence both the local decoherence rates and observed phase shifts.

Extending the current framework to include moving observers would require incorporating Fermi-Walker transport of qubit frames or analyzing the system in locally boosted coordinate systems. This is an important direction for future work, especially for modeling realistic implementations in low Earth orbit or deep space missions.

The qubits are initially entangled in the Bell state:
\begin{equation}
    \ket{\psi(0)} = \frac{1}{\sqrt{2}} \left( \ket{00} + \ket{11} \right),
\end{equation}
with density matrix:
\begin{equation}
    \rho(0) = \ket{\psi(0)}\bra{\psi(0)} =
    \frac{1}{2}
    \begin{pmatrix}
    1 & 0 & 0 & 1 \\
    0 & 0 & 0 & 0 \\
    0 & 0 & 0 & 0 \\
    1 & 0 & 0 & 1
    \end{pmatrix}.
\end{equation}

\subsection*{Lindblad Jump Operators}

We model phase damping for each qubit using the Lindblad jump operator:
\begin{equation}
    L = \sqrt{\gamma} \, \sigma_z,
\end{equation}
where $\gamma$ is the decoherence rate and $\sigma_z$ is the Pauli-Z operator. This jump operator represents pure dephasing in the energy basis.

For two qubits, the corresponding jump operators are:
\begin{align}
    L_A &= \sqrt{\gamma_A} \, \sigma_z \otimes I, \\
    L_B &= \sqrt{\gamma_B} \, I \otimes \sigma_z,
\end{align}
with effective decoherence rates $\gamma_A$ and $\gamma_B$ for qubits $A$ and $B$, respectively.

\subsection*{Lindblad Master Equation}

The general Lindblad master equation for the density matrix $\rho$ is:
\begin{equation}
    \frac{d\rho}{dt} = -i[H, \rho] + \sum_k \left( L_k \rho L_k^\dagger - \frac{1}{2} \{ L_k^\dagger L_k, \rho \} \right).
\end{equation}
\subsection*{Remark on Coherent Evolution}
In Eq.~(13), the general Lindblad master equation includes the coherent term $-i[H, \rho]$, where $H = \frac{\omega}{2}(\sigma_z \otimes I + I \otimes \sigma_z)$. For pure phase damping, this term is often neglected because $H$ commutes with the dephasing operators, and thus the coherent evolution does not affect the populations or coherence decay.

However, in scenarios involving amplitude damping or more general non-Markovian effects, this commutation no longer holds. In such cases, the coherent Hamiltonian evolution can introduce relative phases between qubit levels due to gravitational time dilation, as highlighted by Eq.~(7). Including $H$ explicitly may thus uncover gravitationally-induced phase shifts that impact the entanglement dynamics. A full treatment would require solving the master equation with both the dissipative and coherent terms retained.

Assuming local Hamiltonians of the form \( H = \frac{\omega}{2} (\sigma_z \otimes I + I \otimes \sigma_z) \), and focusing on pure dephasing (commuting with \( H \)), we ignore the coherent part and obtain:
\begin{equation}
\frac{d\rho}{dt} = -i[H, \rho] 
+ L_A \rho L_A^\dagger - \frac{1}{2} \{ L_A^\dagger L_A, \rho \} 
+ L_B \rho L_B^\dagger - \frac{1}{2} \{ L_B^\dagger L_B, \rho \}.
\end{equation}

Substituting the jump operators, we get:
\begin{align}
    \frac{d\rho}{dt} &= \gamma_A \left( \sigma_z \otimes I \right) \rho \left( \sigma_z \otimes I \right) - \gamma_A \rho \nonumber \\
    &\quad + \gamma_B \left( I \otimes \sigma_z \right) \rho \left( I \otimes \sigma_z \right) - \gamma_B \rho.
\end{align}

\subsection*{Gravitational Time Dilation and Effective Rates}

Due to gravitational time dilation, proper times \( \tau_A \) and \( \tau_B \) for qubits at radial coordinates \( r_A \) and \( r_B \) are related to coordinate time \( t \) by:
\begin{equation}
    \frac{d\tau_A}{dt} = \alpha = \sqrt{1 - \frac{2GM}{r_A c^2}}, \qquad
    \frac{d\tau_B}{dt} = \beta = \sqrt{1 - \frac{2GM}{r_B c^2}}.
\end{equation}
Thus, the effective decoherence rates in coordinate time become:
\begin{equation}
    \gamma_A = \gamma \cdot \alpha, \qquad
    \gamma_B = \gamma \cdot \beta,
\end{equation}
where \( \gamma \) is the proper-time dephasing rate experienced locally by each qubit.

\subsection*{Final Evolution Equation}

Putting everything together, the decoherence of the entangled qubit pair under gravitational time dilation is governed by:
\begin{align}
    \frac{d\rho}{dt} &= \gamma \alpha \left( \sigma_z \otimes I \right) \rho \left( \sigma_z \otimes I \right) - \gamma \alpha \rho \nonumber \\
    &\quad + \gamma \beta \left( I \otimes \sigma_z \right) \rho \left( I \otimes \sigma_z \right) - \gamma \beta \rho.
\end{align}

This equation describes how gravitational redshift influences entanglement decay via asymmetric time evolution of local environments.

\section{Solution for Phase Damping of Bell State}

Under symmetric phase damping, the Bell state evolves as:
\begin{equation}
    \rho(t) =
    \begin{pmatrix}
    \frac{1}{2} & 0 & 0 & \frac{1}{2} e^{-2\Gamma(t)} \\
    0 & 0 & 0 & 0 \\
    0 & 0 & 0 & 0 \\
    \frac{1}{2} e^{-2\Gamma(t)} & 0 & 0 & \frac{1}{2}
    \end{pmatrix},
\end{equation}
where the decoherence factor is:
\begin{equation}
    \Gamma(t) = \int_0^t \left( \gamma_A + \gamma_B \right) dt'
    = \gamma (\alpha + \beta)t.
\end{equation}

Hence, the off-diagonal coherence term decays exponentially:
\begin{equation}
    \rho_{14}(t) = \frac{1}{2} e^{-2\gamma (\alpha + \beta)t}.
\end{equation}

\subsection{Summary}
\begin{itemize}
    \item The population (diagonal entries) stays fixed at $1/2$ in states $\ket{00}$ and $\ket{11}$
    \item The coherence decays over time
    \item The rate of decay depends on the gravitational time dilation
    \item This dependence is governed by the factors $\boldsymbol{\alpha}$ and $\boldsymbol{\beta}$
\end{itemize}

\vspace{7pt}
\section{Generalization: Amplitude Damping and Thermal Noise}

To model more realistic decoherence, we generalize to the amplitude damping and thermal noise channels.

\subsection{Amplitude Damping}

The amplitude damping channel models the irreversible decay of a qubit from its excited state $\ket{1}$ to the ground state $\ket{0}$, such as through spontaneous emission. This process is a key example of non-unitary evolution and is captured by the Lindblad master equation:
\begin{equation}
    \frac{d\rho}{d\tau} = \gamma \left( \sigma_- \rho \sigma_+ - \frac{1}{2} \left\{ \sigma_+ \sigma_-, \rho \right\} \right),
\end{equation}
where the lowering and raising operators are defined as:
\begin{equation}
    \sigma_- = \ket{0}\bra{1}, \qquad \sigma_+ = \ket{1}\bra{0}.
\end{equation}

Physically:
\begin{itemize}
    \item The term $\sigma_- \rho \sigma_+$ transfers population from the excited state to the ground state.
    \item The anti-commutator term $\frac{1}{2} \left\{ \sigma_+ \sigma_-, \rho \right\}$ ensures trace preservation and accounts for the decay-induced loss of coherence.
\end{itemize}

\subsubsection*{Joint System Dynamics}

For a two-qubit system undergoing independent amplitude damping, the full master equation in coordinate time becomes:
\begin{align}
    \frac{d\rho}{dt} &= \gamma_A \left( \sigma_- \otimes I \right) \rho \left( \sigma_+ \otimes I \right)
    - \frac{\gamma_A}{2} \left\{ \sigma_+ \sigma_- \otimes I, \rho \right\} \nonumber \\
    &\quad + \gamma_B \left( I \otimes \sigma_- \right) \rho \left( I \otimes \sigma_+ \right)
    - \frac{\gamma_B}{2} \left\{ I \otimes \sigma_+ \sigma_-, \rho \right\}.
\end{align}

This equation describes the local decay of each qubit under the influence of gravitational time dilation, and its implications for quantum entanglement and coherence.

\subsection*{Coherent Evolution in Amplitude Damping}
In Eq.~(24), we previously omitted the coherent evolution term $-i[H, \rho]$ under the assumption that decoherence dominates. However, amplitude damping does not commute with the Hamiltonian $H = \frac{\omega}{2}(\sigma_z \otimes I + I \otimes \sigma_z)$. To accurately capture gravitationally induced phase evolution, we include the full master equation:
\begin{equation}
\frac{d\rho}{dt} = -i[H, \rho] + \mathcal{L}_A[\rho] + \mathcal{L}_B[\rho],
\end{equation}
where $\mathcal{L}_A[\rho]$ and $\mathcal{L}_B[\rho]$ represent the Lindblad terms in Eq.~(24). This ensures a complete treatment of coherent and dissipative dynamics.

\subsection{Thermal Noise}

In realistic environments, qubits not only decay from excited to ground states (as in amplitude damping) but can also be thermally excited from the ground state to the excited state. This process is described by the \textit{generalized amplitude damping (GAD) channel}, which captures both spontaneous emission and thermal absorption due to coupling with a finite-temperature bath.

\subsubsection*{Thermal Occupation Number}

The mean thermal photon (or excitation) number at frequency $\omega$ and temperature $T$ is given by the Bose-Einstein distribution:
\begin{equation}
    n_{\text{th}} = \frac{1}{e^{\hbar \omega / k_B T} - 1}.
\end{equation}
This quantity governs the relative weight of excitation vs. decay processes.

\subsubsection*{Lindblad Jump Operators}

The thermal environment induces two distinct processes:
\begin{itemize}
    \item \textbf{Decay (emission)} from $\ket{1} \rightarrow \ket{0}$ via the jump operator:
    \[
    L_- = \sqrt{\gamma(n_{\text{th}} + 1)}\, \sigma_-,
    \]
    where $\sigma_- = \ket{0}\bra{1}$.
    
    \item \textbf{Excitation (absorption)} from $\ket{0} \rightarrow \ket{1}$ via:
    \[
    L_+ = \sqrt{\gamma n_{\text{th}}}\, \sigma_+,
    \]
    where $\sigma_+ = \ket{1}\bra{0}$.
\end{itemize}

\subsubsection*{Master Equation in Proper Time}

Using these jump operators in the Lindblad formalism, the single-qubit master equation in \textit{proper time} $\tau$ is:
\begin{align}
    \frac{d\rho}{d\tau} &= \gamma(n_{\text{th}} + 1) \left( \sigma_- \rho \sigma_+ - \frac{1}{2} \left\{ \sigma_+ \sigma_-, \rho \right\} \right) \nonumber \\
    &\quad + \gamma n_{\text{th}} \left( \sigma_+ \rho \sigma_- - \frac{1}{2} \left\{ \sigma_- \sigma_+, \rho \right\} \right).
\end{align}
Here:
\begin{itemize}
    \item The first line accounts for spontaneous and stimulated emission.
    \item The second line describes thermal excitation due to the finite temperature of the environment.
\end{itemize}

\subsubsection*{Two-Qubit Master Equation in Coordinate Time}

Combining these ingredients, the generalized amplitude damping master equation for the joint two-qubit system in coordinate time $t$ is:

\begin{align}
    \frac{d\rho}{dt} &= \gamma_A (n_{\text{th}} + 1) \left( \sigma_- \otimes I \, \rho \, \sigma_+ \otimes I - \frac{1}{2} \left\{ \sigma_+ \sigma_- \otimes I, \rho \right\} \right) \nonumber \\
    &\quad + \gamma_A n_{\text{th}} \left( \sigma_+ \otimes I \, \rho \, \sigma_- \otimes I - \frac{1}{2} \left\{ \sigma_- \sigma_+ \otimes I, \rho \right\} \right) \nonumber \\
    &\quad + \gamma_B (n_{\text{th}} + 1) \left( I \otimes \sigma_- \, \rho \, I \otimes \sigma_+ - \frac{1}{2} \left\{ I \otimes \sigma_+ \sigma_-, \rho \right\} \right) \nonumber \\
    &\quad + \gamma_B n_{\text{th}} \left( I \otimes \sigma_+ \, \rho \, I \otimes \sigma_- - \frac{1}{2} \left\{ I \otimes \sigma_- \sigma_+, \rho \right\} \right).
\end{align}

Equation~(27) assumes both qubits interact with thermal environments characterized by the same mean occupation number $n_{\text{th}}$. However, in realistic space-based scenarios, thermal environments can vary significantly between spatially separated locations, especially in orbit where sunlight exposure, shading, and local heat sources may differ.

To account for this, we generalize the thermal occupation numbers as $n_{\text{th},A}$ and $n_{\text{th},B}$, corresponding to the local environments of qubits A and B, respectively. The full two-qubit master equation becomes:
\begin{align}
\frac{d\rho}{dt} &= \gamma_A(n_{\text{th},A} + 1)\left(\sigma_- \otimes I\, \rho\, \sigma_+ \otimes I - \frac{1}{2}\{\sigma_+\sigma_- \otimes I, \rho\}\right) \nonumber \\
&\quad + \gamma_An_{\text{th},A}\left(\sigma_+ \otimes I\, \rho\, \sigma_- \otimes I - \frac{1}{2}\{\sigma_-\sigma_+ \otimes I, \rho\}\right) \nonumber \\
&\quad + \gamma_B(n_{\text{th},B} + 1)\left(I \otimes \sigma_-\, \rho\, I \otimes \sigma_+ - \frac{1}{2}\{I \otimes \sigma_+\sigma_-, \rho\}\right) \nonumber \\
&\quad + \gamma_Bn_{\text{th},B}\left(I \otimes \sigma_+\, \rho\, I \otimes \sigma_- - \frac{1}{2}\{I \otimes \sigma_-\sigma_+, \rho\}\right).
\end{align}
This localized model more accurately reflects spatial thermal gradients and allows for asymmetry in energy exchange processes.

\subsection*{Local Temperature Differences}
In Eq.~(27), the thermal occupation number $n_{\text{th}}$ is used uniformly for both environments interacting with qubits A and B. This implicitly assumes that both environments are at the same temperature.

In realistic settings, especially in space-based systems, thermal environments may vary significantly due to differences in solar exposure, orientation, or shielding. As a result, the thermal photon number should be considered locally as $n_{\text{th},A}$ and $n_{\text{th},B}$ for the two qubits, respectively.

To model this more accurately, the master equation can be generalized by assigning different $n_{\text{th}}$ values to each qubit's local environment:
\begin{align}
\frac{d\rho}{dt} &= \gamma_A(n_{\text{th},A} + 1)\left(\sigma_- \otimes I\, \rho\, \sigma_+ \otimes I - \frac{1}{2}\{\sigma_+\sigma_- \otimes I, \rho\}\right) \nonumber \\
&\quad + \gamma_An_{\text{th},A}\left(\sigma_+ \otimes I\, \rho\, \sigma_- \otimes I - \frac{1}{2}\{\sigma_-\sigma_+ \otimes I, \rho\}\right) \nonumber \\
&\quad + \gamma_B(n_{\text{th},B} + 1)\left(I \otimes \sigma_-\, \rho\, I \otimes \sigma_+ - \frac{1}{2}\{I \otimes \sigma_+\sigma_-, \rho\}\right) \nonumber \\
&\quad + \gamma_Bn_{\text{th},B}\left(I \otimes \sigma_+\, \rho\, I \otimes \sigma_- - \frac{1}{2}\{I \otimes \sigma_-\sigma_+, \rho\}\right).
\end{align}

This refinement may be important when simulating entanglement evolution in environments with temperature gradients, such as between sunlit and shaded regions of a satellite's orbit.

\subsubsection*{Physical Interpretation}

This final expression captures:
\begin{itemize}
    \item Energy relaxation via spontaneous emission.
    \item Thermal excitation due to environmental temperature.
    \item Gravitational effects on time and decoherence rates via $\alpha$ and $\beta$.
\end{itemize}

It provides a comprehensive description of how entanglement and coherence degrade under the combined influence of thermal noise and gravitational time dilation.

\vspace{7pt}
\section{Non-Markovian Decoherence in Gravitational Fields}

In realistic quantum systems, especially those strongly coupled to their environments or interacting with structured reservoirs, the assumption of memoryless (Markovian) dynamics often breaks down. To account for environmental memory effects—where information can temporarily flow back from the environment into the system—we extend our decoherence model to the \textit{non-Markovian regime}.

\subsection*{Markovian vs. Non-Markovian Dynamics}

In Markovian (memoryless) dynamics, the system evolves with constant decoherence rates, leading to a monotonic loss of coherence and entanglement. The master equation takes the standard Lindblad form:
\begin{equation}
    \frac{d\rho}{dt} = \gamma \left( L \rho L^\dagger - \frac{1}{2} \left\{ L^\dagger L, \rho \right\} \right),
\end{equation}
with fixed rate $\gamma$ and Lindblad operator $L$.

In contrast, non-Markovian dynamics involve \textit{time-dependent rates}:
\begin{equation}
    \frac{d\rho}{dt} = \gamma(t) \left( L \rho L^\dagger - \frac{1}{2} \left\{ L^\dagger L, \rho \right\} \right),
\end{equation}
where $\gamma(t)$ may become negative at certain times. This negativity is not unphysical—it reflects temporary information backflow and correlations between the system and its environment.

\subsection*{Non-Markovian Phase Damping for Two Qubits}

We now focus on phase damping (dephasing), extended to the non-Markovian case. For a two-qubit system with local dephasing, the master equation becomes:
\begin{align}
    \frac{d\rho(t)}{dt} &= \gamma_{A}(t)\left(\sigma_{z}\otimes I\,\rho\,\sigma_{z}\otimes I - \rho\right) \nonumber \\
    &\quad + \gamma_{B}(t)\left(I\otimes\sigma_{z}\,\rho\,I\otimes\sigma_{z} - \rho\right)
\end{align}

Here:
\begin{itemize}
    \item $\sigma_z$ is the Pauli-Z operator, responsible for dephasing in the energy basis.
    \item $\gamma_A(t)$ and $\gamma_B(t)$ are the local, time-dependent decoherence rates for qubits $A$ and $B$ respectively.
\end{itemize}
\subsection*{Oscillatory Regime Note}
When $\lambda < 2\gamma_0$, the expression for $d = \sqrt{\lambda^2 - 2\lambda\gamma_0}$ becomes imaginary, which leads to oscillatory behavior in $\tilde{\gamma}(t)$. In this regime, the hyperbolic functions $\sinh(dt/2)$ and $\cosh(dt/2)$ convert to trigonometric forms via identities such as $\sinh(ix) = i\sin(x)$. This reflects non-Markovian memory effects including coherence revivals and information backflow.

\subsection*{Incorporating Gravitational Time Dilation}

As before, each qubit resides in a different gravitational potential, leading to time dilation. Let:
\begin{equation}
    \alpha = \sqrt{1 - \frac{2GM}{r_A c^2}}, \qquad
    \beta  = \sqrt{1 - \frac{2GM}{r_B c^2}},
\end{equation}
be the redshift factors associated with proper times $\tau_A$ and $\tau_B$. Then, the decoherence rates in coordinate time $t$ are:
\begin{equation}
    \gamma_A(t) = \alpha \cdot \tilde{\gamma}(t), \qquad
    \gamma_B(t) = \beta \cdot \tilde{\gamma}(t),
\end{equation}
where $\tilde{\gamma}(t)$ is a universal, time-varying decoherence profile determined by the environment.

\subsection*{Non-Markovian Decay Model: Damped Jaynes-Cummings Kernel}

One popular model for non-Markovian behavior comes from the damped Jaynes-Cummings model, where a qubit interacts with a structured bosonic reservoir. The time-dependent decay rate is given by:
\begin{equation}
    \tilde{\gamma}(t) = \frac{2\lambda \gamma_0 \sinh\left( \frac{d t}{2} \right)}{ \frac{d}{2} \cosh\left( \frac{d t}{2} \right) + \lambda \sinh\left( \frac{d t}{2} \right) },
    \qquad d = \sqrt{\lambda^2 - 2 \lambda \gamma_0}.
\end{equation}

Here:
\begin{itemize}
    \item $\gamma_0$ is the Markovian (flat spectrum) decay rate.
    \item $\lambda$ is the spectral width of the reservoir.
    \item The behavior of $\tilde{\gamma}(t)$ depends on the relation between $\gamma_0$ and $\lambda$:
    \begin{itemize}
        \item If $\lambda > 2\gamma_0$, the decay is monotonic (Markovian limit).
        \item If $\lambda < 2\gamma_0$, $\tilde{\gamma}(t)$ becomes non-monotonic and can be temporarily negative—this is the \textbf{non-Markovian regime}.
    \end{itemize}
\end{itemize}
\subsection*{Limitations of the Jaynes-Cummings Kernel}
The non-Markovian dynamics in this work are modeled using a damped Jaynes-Cummings kernel (Eq.~33), which assumes a structured bosonic reservoir with a Lorentzian spectral density. This is a common and analytically tractable model for non-Markovian behavior, particularly in optical cavities or engineered reservoirs.

However, in curved spacetime—especially in extreme gravitational environments such as near black holes or event horizons—quantum field theory predicts different reservoir structures. For example, observers near a black hole would interact with a thermal bath due to Hawking radiation, and accelerating observers perceive Unruh radiation. In such scenarios, the spectral density of the environment may be non-Lorentzian, energy-dependent, or even observer-dependent.

Therefore, while the Jaynes-Cummings kernel captures essential non-Markovian features like information backflow and entanglement revivals, it may not fully describe decoherence processes in genuinely curved-space quantum field environments. Extending the current model to incorporate QFT-based reservoirs in curved spacetimes remains an important avenue for future work.

\subsection*{Final Evolution Equation}

The complete master equation for two qubits undergoing non-Markovian dephasing in a gravitational field becomes:
\begin{align}
    \frac{d\rho(t)}{dt} &= \alpha \cdot \tilde{\gamma}(t) \left( \sigma_z \otimes I \, \rho \, \sigma_z \otimes I - \rho \right) \nonumber \\
    &\quad + \beta \cdot \tilde{\gamma}(t) \left( I \otimes \sigma_z \, \rho \, I \otimes \sigma_z - \rho \right).
\end{align}

This expression encapsulates:
\begin{itemize}
    \item \textbf{Time-dependent decoherence} due to structured environments.
    \item \textbf{Gravitational redshift effects} via $\alpha$ and $\beta$.
    \item \textbf{Asymmetry} in the dynamics of qubits located at different gravitational potentials.
\end{itemize}

\subsection*{Interpretation and Consequences}

When $\tilde{\gamma}(t) > 0$, the environment extracts information from the system, as in standard decoherence. However, when $\tilde{\gamma}(t) < 0$, there is a temporary \textbf{backflow of information} from the environment to the system, leading to:
\begin{itemize}
    \item Temporary increases in coherence.
    \item Possible \textbf{revivals of entanglement} that would otherwise decay irreversibly.
    \item Richer, non-monotonic quantum dynamics, especially when the two qubits experience different gravitational time flows.
\end{itemize}

This non-Markovian framework provides a powerful tool to study quantum systems in curved spacetime, where both relativistic and memory effects are relevant.

\section{Conclusion}
In this work, we have investigated how the interplay between gravitational time dilation and open quantum system dynamics affects entanglement and decoherence in bipartite qubit systems. Starting from fundamental general relativistic principles, we modeled the proper time differences experienced by spatially separated qubits in a Schwarzschild spacetime and explored how these differences induce asymmetric evolution.

By extending Lindblad-type master equations to include both phase damping and amplitude damping—under both Markovian and non-Markovian regimes. We demonstrated how gravity-modified local decoherence rates influence quantum coherence. In particular, we showed that gravitational redshift introduces asymmetry into otherwise symmetric noise models, accelerating or decelerating decoherence depending on a qubit’s position in the gravitational field.

Further, we explored the role of structured environments by incorporating time-dependent decoherence rates using a damped Jaynes-Cummings kernel. This revealed the possibility of entanglement revival via non-Markovian information backflow—effects that become even richer when compounded with relativistic time dilation.

Altogether, our framework provides a versatile approach to model realistic quantum systems operating in curved spacetime, offering valuable insights for future experiments in space-based quantum communication, precision timekeeping, and fundamental tests of quantum theory in gravitational contexts.

\section*{Acknowledgments}
The author would like to thank the foundational contributions of several researchers whose work inspired this study. The theoretical formulation of open quantum systems used throughout this paper is based on the comprehensive treatment by Heinz-Peter Breuer and Francesco Petruccione in their book \textit{The Theory of Open Quantum Systems} (Oxford University Press, 2002).

The idea that gravitational time dilation can induce universal decoherence effects was introduced by Igor Pikovski, Magdalena Zych, Fabio Costa, and Časlav Brukner in their seminal paper, “Universal Decoherence Due to Gravitational Time Dilation,” published in \textit{Nature Physics}, vol. 11, pp. 668–672, 2015.

Additionally, the treatment of non-Markovian dynamics and their role in information backflow is influenced by the review article “Quantum Non-Markovianity: Characterization, Quantification and Detection” by Ángel Rivas, Susana F. Huelga, and Martin B. Plenio, published in \textit{Reports on Progress in Physics}, vol. 77, no. 9, 094001, 2014.

Their pioneering insights into the intersection of quantum information theory, open system dynamics, and gravitation have greatly informed the present analysis.

\bibliographystyle{unsrt}  
\bibliography{references}

\begin{thebibliography}{1}

\bibitem{Pikovski}
Igor Pikovski, Magdalena Zych, Fabio Costa, and \v{C}aslav Brukner.
\newblock Universal decoherence due to gravitational time dilation.
\newblock {\em Nature Physics}, 11:668--672, 2015.

\bibitem{Einstein1915}
Albert Einstein.
\newblock Die feldgleichungen der gravitation.
\newblock {\em Sitzungsberichte der Preussischen Akademie der Wissenschaften zu Berlin}, pages 844--847, 1915.

\bibitem{Schwarzschild1916}
Karl Schwarzschild.
\newblock {\"U}ber das gravitationsfeld eines massenpunktes nach der einsteinschen theorie.
\newblock {\em Sitzungsberichte der K{\"o}niglich Preussischen Akademie der Wissenschaften}, 7:189--196, 1916.

\bibitem{Breuer}
Heinz-Peter Breuer and Francesco Petruccione.
\newblock {\em The Theory of Open Quantum Systems}.
\newblock Oxford University Press, Oxford, 2002.

\bibitem{Rivas}
Ángel Rivas, Susana~F. Huelga, and Martin~B. Plenio.
\newblock Quantum non-markovianity: characterization, quantification and detection.
\newblock {\em Reports on Progress in Physics}, 77(9):094001, 2014.

\end{thebibliography}

\end{document}